\documentclass[aip,jcp,reprint]{revtex4-1}
\usepackage{graphicx}
\usepackage{amsmath}
\usepackage{bm}
\usepackage{amssymb} 
\usepackage[usenames, dvipsnames]{color}

\begin{document}

\title{Electrolytes confined between polarizable surfaces in slit pores with anisotropic permittivity tensor}

\author{Alexandre P. dos Santos}
\email{alexandre.pereira@ufrgs.br}
\affiliation{Instituto de F\'isica, Universidade Federal do Rio Grande do Sul, Caixa Postal 15051, CEP 91501-970, Porto Alegre, RS, Brazil.}

\author{Yan Levin}
\email{levin@if.ufrgs.br}
\affiliation{Instituto de F\'isica, Universidade Federal do Rio Grande do Sul, Caixa Postal 15051, CEP 91501-970, Porto Alegre, RS, Brazil.}

\begin{abstract}

We present a method that enables efficient simulations of coarse-grained electrolyte solutions inside a narrow slit pore with an anisotropic dielectric permittivity tensor. The electrostatic equations for polarizable surfaces are solved using a 2D periodic Green's function method combined with a slab-corrected anisotropic 3D Ewald summation. We apply this approach in Monte Carlo simulations to study 1:1 electrolytes confined between both polarizable dielectric and metallic surfaces. Our results show that dielectric anisotropy aggressively reshapes the double-layer structure. While a coordinate stretching transformation demonstrates that individual ion-image interactions depend strictly on the bulk-like parallel permittivity, the suppression of the perpendicular permittivity dramatically amplifies direct in-plane ion-ion correlations. Under strong anisotropy, these lateral correlations dominate the thermodynamics completely, rendering the structural profiles of mutually opposing dielectric and metallic boundaries practically identical by forcing the smaller cations directly into the contact plane of the larger anions.

\end{abstract}
 
\maketitle

\section{Introduction}
This paper is dedicated to the memory of Rudi Podgornik, whose recent passing is a profound loss to the scientific community. Rudi was one of the pioneers of statistical mechanics of Coulomb fluids, macromolecular interactions, and soft matter physics. His profound insights into electrostatic correlations, charge regulation, and fluctuation-induced attraction have helped us to understand how charged entities interact in complex, condensed-phase environments. His legacy of intellectual rigor and deep curiosity, continues to inspire the field of colloid and interface science. It is therefore highly fitting that the work presented here, which explores how low-dimensional correlation zones emerge from anisotropic dielectric boundaries, builds directly upon the rich traditions of theoretical physics that Rudi championed throughout his remarkable career.

Electrolyte solutions under nanoscale confinement exhibit physical properties that can differ markedly from their bulk counterparts~\cite{doi:10.1021/acs.langmuir.5c05917,doi:10.1021/acs.jpclett.0c03219,doi:10.1021/acsnano.0c03173,FlZh22,D3FD00062A,doi:10.1021/acs.langmuir.0c03116,10.1063/1.4962198,doi:10.1021/acs.jpclett.4c01698,doi:10.1021/acs.chemrev.2c00155,PhysRevLett.131.118201,doi:10.1021/acs.jpcc.7b11093,https://doi.org/10.1002/adfm.202314288,https://doi.org/10.1002/adma.202400508,PhysRevLett.116.225901,doi:10.1021/jp4098638,10.1063/1.3490666,doi:10.1021/jacs.5c04186,doi:10.1021/acs.nanolett.3c01171,doi:10.1021/acs.jpclett.8b02473,PhysRevResearch.5.043174,Telles_2026,C5NR00508F,Levin_2002}. In particular, when the characteristic confinement length approaches the nanometer scale, the dielectric response of the medium becomes highly nontrivial, independent of geometry~\cite{PhysRevLett.117.048001,doi:10.1021/acs.chemrev.2c00155,10.1063/5.0032879,doi:10.1021/acs.jpcb.9b09269}. Recent experimental measurements and molecular simulations have shown that, for confinements below approximately $1$nm, the dielectric permittivity becomes anisotropic, with the component perpendicular to the confining surfaces significantly reduced compared to the parallel components~\cite{10.2138/rmg.2026.91B.02,doi:10.1021/acs.jpcb.9b09269,doi:10.1126/science.aat4191,PhysRevLett.117.048001,10.1063/1.1845431,PhysRevE.102.022803,10.1063/5.0315836,qp52-s5d5,doi:10.1021/acs.jctc.5c00354,PhysRevMaterials.5.024008,doi:10.1021/acs.jpclett.1c00447,doi:10.1021/acs.langmuir.6b00791,doi:10.1021/acsnano.1c08512,doi:10.1021/jz401108n,doi:10.1021/acs.jpclett.4c01242,10.1063/5.0038359}.

This dielectric anisotropy has important consequences for electrostatic interactions, ion distributions, and thermodynamic properties in confined systems, yet it remains insufficiently incorporated into continuum and coarse-grained theoretical descriptions.
A common framework for studying confined electrolytes relies on the solution of the Poisson equation subject to the appropriate boundary conditions imposed by dielectric interfaces. For isotropic media, this problem has been extensively investigated, including for systems bounded by polarizable surfaces~\cite{10.1063/1.4921221,PhysRevLett.111.106102,10.1063/5.0314749,doi:10.1021/jp503224w,10.1063/1.4935704,10.1063/5.0233310,PhysRevE.69.046702,GAN2015317,10.1063/1.3376011,10.1063/1.4997420,e15114569,10.1063/1.3021064,10.1063/1.2790428}. However, extending these approaches to anisotropic dielectric environments introduces additional complexity, particularly when permittivity must be described by a tensor rather than a scalar quantity.

Efficient and accurate methods to account for such anisotropy are therefore needed to bridge the gap between microscopic observations and mesoscopic modeling.
In this work, we present a method to simulate coarse-grained electrolyte solutions confined between polarizable surfaces in the presence of dielectric anisotropy. The anisotropic response of the medium is represented by a diagonal permittivity tensor, allowing for distinct parallel and perpendicular components. By employing the periodic Green’s function method~\cite{10.1063/1.4997420,10.1063/1.4945560}, we solve the Poisson equation for polarizable boundaries through a simple transformation of coordinates in the direction perpendicular to the confining surfaces, together with the introduction of an effective dielectric constant. This approach preserves computational efficiency while incorporating the essential physics of dielectric anisotropy.
We apply the proposed method within Monte Carlo simulations to investigate a symmetric 1:1 electrolyte confined between polarizable surfaces. Our results demonstrate that dielectric anisotropy significantly alters ionic distributions near the interfaces compared to the isotropic case, highlighting the importance of properly accounting for directional dielectric response in strongly confined systems.

\section{Calculations}

Consider a particle with charge $q_i$ located at ${\bf r}_i=(x_i,y_i,z_i)$, confined between two planar surfaces located at $z=0$ and $z=L$. The simulation box is limited in $x$ and $y$ directions by length $L_{xy}$ and it is replicated infinitely in these directions. The permittivity in the confining region, $0<z<L$, is represented by the tensor $\boldsymbol{\varepsilon} = \mathrm{diag}(\varepsilon_{\parallel}, \varepsilon_{\parallel}, \varepsilon_{\perp})$, while outside it is uniform, $\varepsilon_o$.

Using the idea of periodic Green functions developed in Ref.~\cite{10.1063/1.4997420},
the Green's function in confining region obeys the Poisson equation,
\begin{equation}
\varepsilon_{\parallel} \dfrac{\partial^2 G}{\partial x^2}+\varepsilon_{\parallel} \dfrac{\partial^2 G}{\partial y^2}+\varepsilon_{\perp} \dfrac{\partial^2 G}{\partial z^2}=-4 \pi q_i \sum_{{\bf n}=-\infty}^\infty \delta({\bf r}-{\bf r}_i+L_{xy}{\bf n}) \ ,
\label{poisson}
\end{equation}
where ${\bf n}=(n_x,n_y)$, and $n_x$ and $n_y$ are integers.
The periodicity allows us to define
\begin{equation}
G({\bf r},{\bf r}_i)=\frac{1}{L_{xy}^2}\sum_{{\bf n}=-\infty}^{\infty}g_{\bf m}(z,z_i)\exp{[i{\bf m}\cdot ({\bm \rho}-{\bm \rho}_i)]} \ ,
\end{equation}
where ${\bf m}=(2\pi n_x/L_{xy},2\pi n_y/L_{xy})$ and ${\bm \rho}=(x,y)$.
Inserting this into Eq.~\eqref{poisson}, we obtain
\begin{equation}
\varepsilon_{\perp}\frac{\partial^2}{\partial z^2}g_{\bf m}(z,z_i)-\varepsilon_{||} m^2 g_{\bf m}(z,z_i) = -4 \pi q_i \delta(z-z_i) \ ,
\end{equation}
where $m=\dfrac{2\pi}{L_{xy}}\sqrt{n_x^2+n_y^2}$.
Changing variable $\tilde{z}=\sqrt{\frac{\varepsilon_{\parallel}}{\varepsilon_{\perp}}}z$ and using the geometric mean permittivity $\varepsilon_{eff}=\sqrt{\varepsilon_{\parallel}\varepsilon_{\perp}}$, this equation can be rewritten as:
\begin{equation}
\frac{\partial^2}{\partial \tilde{z}^2}g_{\bf m}(\tilde{z},\tilde{z}_i)-m^2 g_{\bf m}(\tilde{z},\tilde{z}_i) = -\frac{4 \pi q_i}{\varepsilon_{eff}}\delta(\tilde{z}-\tilde{z}_i) \ .
\end{equation}

The solution for the Green function and the polarization contribution to the electrostatic energy can then be directly extracted from Ref.~\cite{10.1063/1.4997420}, with the proper change of variables and definitions, $\tilde{z}=\sqrt{\dfrac{\varepsilon_{\parallel}}{\varepsilon_{\perp}}}z$, $\tilde{L}=\sqrt{\dfrac{\varepsilon_{\parallel}}{\varepsilon_{\perp}}}L$ and $\gamma=\dfrac{\varepsilon_{eff}-\varepsilon_o}{\varepsilon_{eff}+\varepsilon_o}$,
\begin{equation}
\begin{split}
&U_p=\frac{\pi}{\varepsilon_{eff} L_{xy}^2}\sum_{{\bf n}'=-\infty}^{\infty} \frac{\gamma}{m (1-\gamma^2 e^{-2 m \tilde{L}})}\{ f_1({\bf m})^2 +
f_2({\bf m})^2 + \\
& +e^{-2 m \tilde{L}} \left(f_3({\bf m})^2 +f_4({\bf m})^2 \right) + 2\gamma e^{-2 m \tilde{L}}[f_3({\bf m})f_1({\bf m}) + \\
& +f_2({\bf m})f_4({\bf m})]\} \ ,
\end{split}   
\end{equation}
where the prime symbol over ${\bf n}$ means that the term $n_x=n_y=0$ is not included in the summation.  The functions are 
\begin{equation}
f_1({\bf m})=\sum_{i=1}^{N} q_i \cos\left( {\bf m}\cdot{\bm \rho}_i \right)e^{-m\tilde{z}_i} \ ,   
\end{equation}
\begin{equation}
f_2({\bf m})=\sum_{i=1}^{N} q_i \sin\left( {\bf m}\cdot{\bm \rho}_i \right)e^{-m\tilde{z}_i} \ ,
\end{equation}
\begin{equation}
f_3({\bf m})=\sum_{i=1}^{N} q_i \cos\left( {\bf m}\cdot{\bm \rho}_i \right)e^{m\tilde{z}_i} \ ,
\end{equation}   
\begin{equation}
f_4({\bf m})=\sum_{i=1}^{N} q_i \sin\left( {\bf m}\cdot{\bm \rho}_i \right)e^{m\tilde{z}_i} \ .
\end{equation}
These functions can be efficiently updated for each MC move, since they involve only the individual particle  coordinates.
The $n_x=n_y=0$ contribution to the sum is zero for $-1<\gamma<1$. For $\gamma=-1$ (metallic cases) it is
\begin{equation}
\begin{split} 
U_{(-1)}=-\frac{2\pi}{L_{xy}^2} \left[ \frac{\tilde{M}_z^2}{\tilde{L}} - Q_t \tilde{M}_z \right] \ .
\end{split}
\end{equation}

The total electrostatic energy is then
\begin{equation}
U=U_{ES}+U_p+U_{\gamma} \,
\end{equation}
where $U_{\gamma}$ is non-zero only for metal surfaces with $\gamma=-1$.

The direct Coulomb interaction energy, $U_{ES}$, can be efficiently calculated using a 3D Ewald summation corrected for the slab geometry~\cite{10.1063/1.4945560}. To implement this, we embed the confining slit region into a simulation box artificially replicated in all three dimensions. The box lengths are $L_{xy}$ in the $x$ and $y$ directions, and $L_z$ in the $z$ direction, where we enforce $L_z > 2L_{xy}$ and $L_z > 2L$ to accurately apply the slab correction. In this 3D replicated space, the permittivity tensor remains $\boldsymbol{\epsilon} = \text{diag}(\epsilon_{||}, \epsilon_{||}, \epsilon_{\perp})$ throughout the periodic domain, and the 3D periodic Green's function obeys:
\begin{equation}
\varepsilon_{\parallel} \dfrac{\partial^2 G}{\partial x^2}+\varepsilon_{\parallel} \dfrac{\partial^2 G}{\partial y^2}+\varepsilon_{\perp} \dfrac{\partial^2 G}{\partial z^2}=-4 \pi q_i \sum_{{\bf n}=-\infty}^\infty \delta({\bf r}-{\bf r}_i+{\bf p}) \ ,
\label{poissonew}
\end{equation}
where ${\bf p}=(n_x L_{xy},n_y L_{xy},n_z L_z)$, $n_x$, $n_y$ and $n_z$ are integers.
We can rewrite the equation using again the auxiliary variable $\tilde{z}=\sqrt{\frac{\varepsilon_{\parallel}}{\varepsilon_{\perp}}}z$ and the definition $\varepsilon_{eff}=\sqrt{\varepsilon_{\parallel}\varepsilon_{\perp}}$,
\begin{equation}
\dfrac{\partial^2 G}{\partial x^2}+\dfrac{\partial^2 G}{\partial y^2}+\dfrac{\partial^2 G}{\partial \tilde{z}^2}=-\frac{4 \pi q_i}{\varepsilon_{eff}} \sum_{{\bf n}=-\infty}^\infty \delta({\bf \tilde{r}}-{\bf \tilde{r}}_i+{\bf \tilde{p}}) \ ,
\label{poisson2}
\end{equation}
where ${\bf \tilde{r}}=(x,y,\tilde{z})$, ${\bf \tilde{r}}_i=(x_i,y_i,\tilde{z}_i)$ and ${\bf \tilde{p}}=(n_x L_{xy},n_y L_{xy},n_z \tilde{L}_z)$.

The electrostatic energy of a system with $N$ charges can be obtained using the Ewald summation with correction for slab geometry~\cite{10.1063/1.4945560}, with the proper change of variables
\begin{eqnarray}\label{ener}
U_{ES}=\sum_{{\bf n}'=-\infty}^{\infty}\frac{2\pi}{\varepsilon_{eff}\tilde{V} |\tilde{\pmb k}|^2}\exp{[-
\frac{|\tilde{\pmb k}|^2}{4\kappa_e^2}]}[A(\tilde{\pmb k})^2+B(\tilde{\pmb k})^2] + \nonumber \\
\frac{2\pi}{\varepsilon_{eff}\tilde{V}}[\tilde{M}_z^2-Q_t\tilde{G}_z]
+ \hspace{1cm} \nonumber \\
\dfrac{1}{2}\sum_{i \ne j}^N q_iq_j\frac{\text{erfc}(\kappa_e|\tilde{\pmb r}_i-\tilde{\pmb r}_j|)}{\varepsilon_{eff}
|\tilde{\pmb r}_i-\tilde{\pmb r}_j|} \ , \hspace{1cm}
\end{eqnarray}
where
\begin{eqnarray}
A(\tilde{\pmb k})=\sum_{i=1}^N q_i\text{cos}(\tilde{\pmb k}\cdot\tilde{\pmb r}_i) \ \ , \ \  
B(\tilde{\pmb k})=-\sum_{i=1}^N q_i\text{sin}(\tilde{\pmb k}\cdot\tilde{\pmb r}_i) \ \ , \ \ \nonumber \\
\tilde{M}_z=\sum_{i=1}^N q_i \tilde{z}_i \ \ , \ \
Q_t=\sum_{i=1}^N q_i \ \ , \ \
\tilde{G}_z=\sum_{i=1}^N q_i \tilde{z}_i^2 \ .
\end{eqnarray}
The k-vector is $\tilde{\pmb k}=(2\pi n_x/L_{xy},2\pi n_y/L_{xy},2\pi n_z /\tilde{L}_z)$, while the volume is $\tilde{V}=L_{xy}^2\tilde{L}_z$.  Note that this expression remains valid even for charge non-neutral systems~\cite{10.1063/1.4945560}.

\section{Model and Monte Carlos simulations}

\begin{figure}[h]
\includegraphics[scale=0.5]{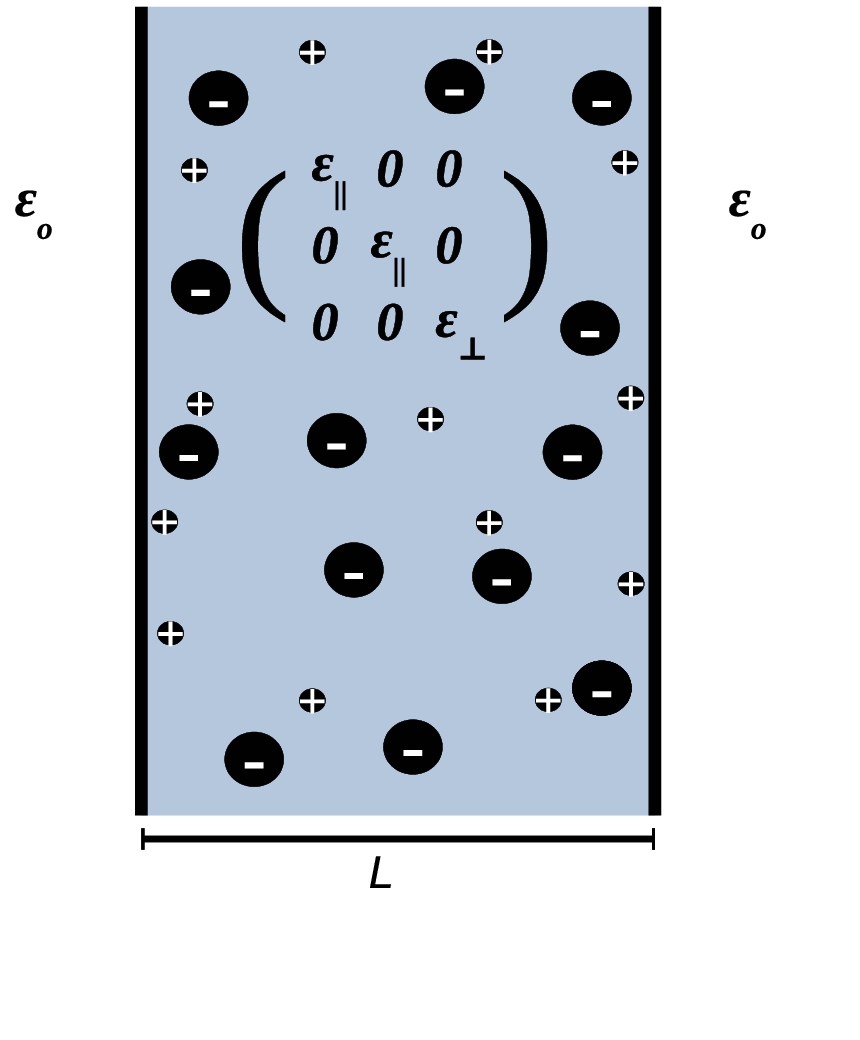}
\caption{Illustration of the system. The cations and anions are confined between two planar polarizable surfaces, separated by distance $L$. Outside, the dielectric constant is $\varepsilon_o$ while inside it is given by the tensor $\mathrm{diag}(\varepsilon_{\parallel}, \varepsilon_{\parallel}, \varepsilon_{\perp})$.}
\label{fig:ilul}
\end{figure}

To test our approach, we study a 1:1 size-asymmetric electrolyte solution confined within a slit pore bounded by two planar polarizable surfaces located at $z = \pm 0.5\text{ nm}$, as illustrated schematically in Fig.~\ref{fig:ilul}. The system consists of $N_+ = 100$ cations and $N_- = 100$ anions carrying elementary charges $+e$ and $-e$, respectively. In addition to long-range electrostatic forces, the ions are subject to hard-core excluded volume interactions that prevent them from overlapping with one another or penetrating the confining boundaries. To capture steric asymmetry, the cations and anions are assigned effective ionic diameters of $d_+ = 0.3\text{ nm}$ and $d_- = 0.6\text{ nm}$, respectively. The total separation between the two surfaces is $L = 1\text{ nm}$, while the lateral dimensions of the simulation box are set to $L_{xy} = 4\text{ nm}$. 

Canonical Monte Carlo (MC) simulations are performed using the standard Metropolis algorithm for various dielectric pore configurations. For each system, $10^6$ MC steps are used to achieve thermal equilibrium. Following equilibration, production runs are carried out where a total of $10^4$ statistically independent configurations are stored for further analysis, sampling the system every $10^4$ MC steps.

\section{Results and Discussion}

To understand the physical mechanisms driving the structural variations in the ionic profiles, it is instructive to first isolate how the dielectric anisotropy alters both the individual ion-wall interactions and the inter-ionic pair potentials. This can be understood analytically by applying a coordinate stretching transformation along the confinement axis, $\tilde{z} = \sqrt{\epsilon_{||}/\epsilon_{\perp}}z$, while keeping the physical charge $q$ unscaled. In this transformed coordinate frame, the anisotropic Poisson equation maps directly onto an isotropic form characterized by an effective isotropic dielectric constant, $\epsilon_{\text{eff}} = \sqrt{\epsilon_{||}\epsilon_{\perp}}$.

When this mapping is applied to an isolated ion at a distance $d$ from a planar boundary, the flat interface simply shifts to $\tilde{d} = \sqrt{\epsilon_{||}/\epsilon_{\perp}}d$ in the stretched coordinate frame. Evaluating the interaction of the charge with its corresponding image charge in a medium of permittivity $\epsilon_{\text{eff}}$ yields a self-image potential energy of:
\begin{equation}
U_{\text{self}}(d) = \frac{\gamma q^2}{4 \epsilon_{\text{eff}} \tilde{d}} = \frac{\gamma q^2}{4 \left(\sqrt{\epsilon_{||}\epsilon_{\perp}}\right) \left(\sqrt{\epsilon_{||}/\epsilon_{\perp}}d\right)} = \frac{\gamma q^2}{4 \epsilon_{||} d},
\end{equation}
where the dielectric contrast factor is defined as $\gamma = (\epsilon_{\text{eff}} - \epsilon_o)/(\epsilon_{\text{eff}} + \epsilon_o)$. This result reveals a key feature of the model: the direct spatial stretching contribution of $\epsilon_{\perp}$ in the denominator cancels out completely, leaving the potential inversely proportional to the bulk-like parallel permittivity $\epsilon_{||}$. Consequently, the perpendicular permittivity component affects the baseline single-ion image force strictly through the boundary contrast factor $\gamma$. Notably, for ideal metallic boundaries ($\epsilon_o = \infty$), the contrast factor becomes a constant $\gamma = -1$, rendering the individual image well completely independent of $\epsilon_{\perp}$ and quantitatively identical to that of a homogeneous bulk medium.

The dramatic structural re-organization of the double layer for small  $\epsilon_{\perp}$ is mediated entirely by the modification of the direct ion-ion Coulomb interactions. In the stretched coordinate frame, the pair potential between two ions scales as $q_i q_j / (\epsilon_{\text{eff}} \tilde{r}_{ij})$, which explicitly expands to $q_i q_j / [ \sqrt{\epsilon_{||}\epsilon_{\perp}}\sqrt{\Delta x^2 + \Delta y^2 + (\epsilon_{||}/\epsilon_{\perp})\Delta z^2} ]$. Crucially, for any two ions residing within the same transverse layer ($\Delta z = 0$), the stretched spatial component vanishes identically. The resulting in-plane interaction potential reduces to:
\begin{equation}
U_{ij}\Big|_{z_i = z_j} = \frac{q_i q_j}{\sqrt{\epsilon_{||}\epsilon_{\perp}}\rho_{ij}},
\end{equation}
where $\rho_{ij} = \sqrt{\Delta x^2 + \Delta y^2}$ is the lateral separation distance. As $\epsilon_{\perp}$ is lowered from the bulk value of $78.54$ down to $4$, the screening prefactor $1/\sqrt{\epsilon_{||}\epsilon_{\perp}}$ increases substantially relative to the bulk-like $1/\epsilon_{||}$. This leads to very strong correlations between cations and anions, forcing them to move into the same $z$-plane.

\subsection{Polarizable Dielectric Surfaces}

\begin{figure}[h]
\includegraphics[scale=0.3]{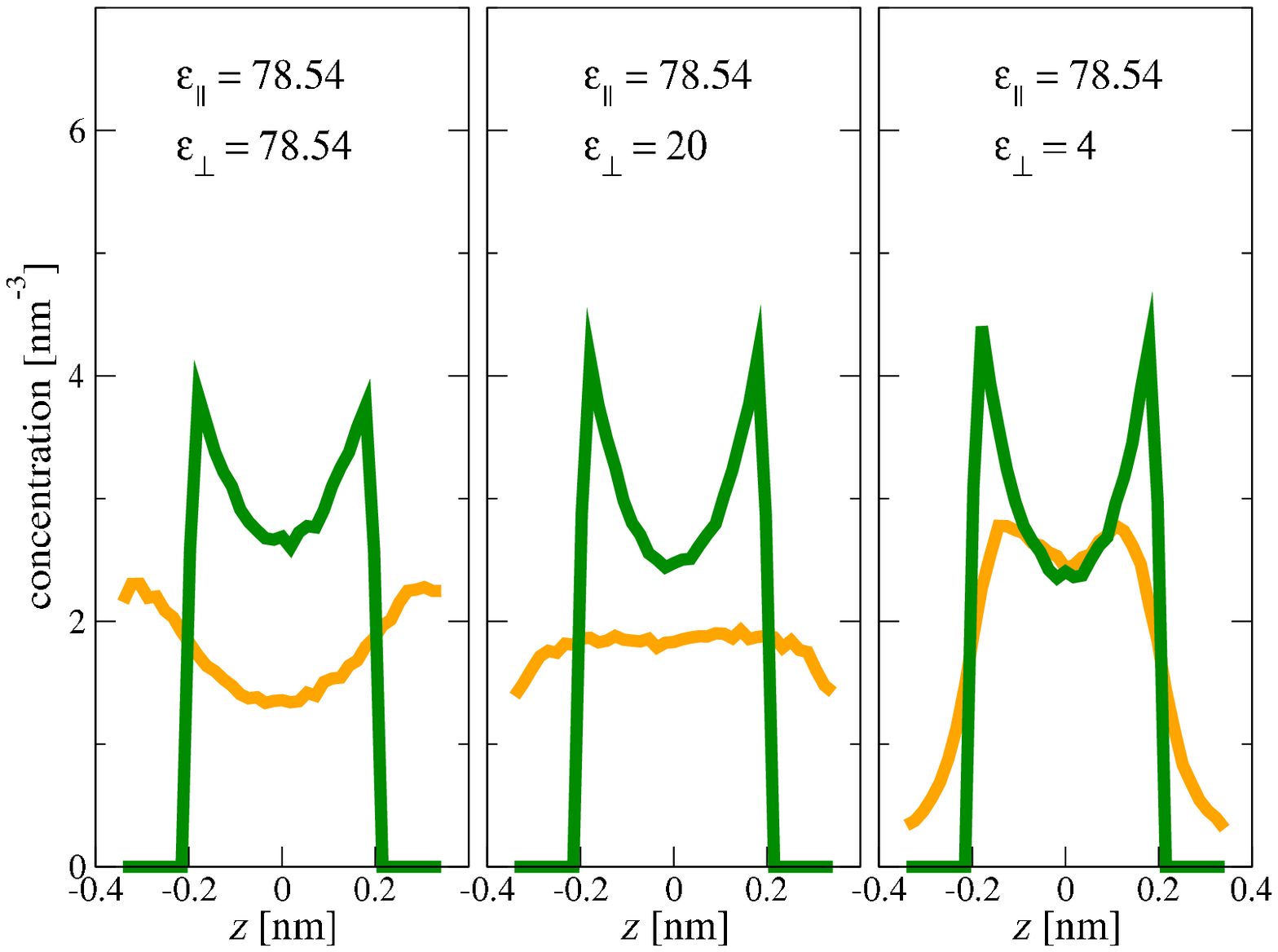}
\caption{Concentration profiles of cations (orange) and anions (green) for three different dielectric scenarios explained in legends, for dielectric confining surfaces.}
\label{fig:profs1}
\end{figure}

We first consider the case of polarizable dielectric slit pores, where the external medium possesses a low dielectric constant ($\epsilon_o = 4$), corresponding to a positive dielectric contrast factor ($\gamma > 0$) that induces image repulsion, see Fig.~\ref{fig:ilul}. Fig.~\ref{fig:profs1} displays the ionic concentration profiles for an size asymmetric 1:1 electrolyte mixture, where the large anions have a hard-core radius twice as large as the cations ($d_- = 0.6\text{ nm}$ and $d_+ = 0.3\text{ nm}$), confined within a slit pore with walls located at $\pm 0.5\text{ nm}$. Due to these steric constraints, the physical motion of the ions is strictly bounded, setting distinct contact planes for each species.

As seen in Fig.~\ref{fig:profs1}, when the medium is isotropic ($\epsilon_{||} = \epsilon_{\perp} = 78.54$), the ions exhibit standard electric double-layer structuring, with the large anions packing against their contact plane. As the perpendicular permittivity $\epsilon_{\perp}$ is systematically reduced to $20$ and $4$ while holding $\epsilon_{||}$ constant, we observe a profound rearrangement. Because the individual image repulsion barrier $U_{\text{self}}(d)$ depends only on $\epsilon_{||}$, the physical driving force for this reorganization is not an enhanced wall repulsion. Rather, it is the intense amplification of the in-plane pair interactions ($1/\sqrt{\epsilon_{||}\epsilon_{\perp}}$).

The sterically crowded anions are pinned at their contact plane. Due to the enhanced in-plane attraction, the smaller cations are powerfully drawn out of their own preferred bulk-like positions and lock directly into the exact same $z$-plane occupied by the anions. This cooperative in-plane alignment optimizes local charge neutrality within the two-dimensional layer and maximizes direct Coulombic screening. 

\begin{figure}[h]
\includegraphics[scale=0.3]{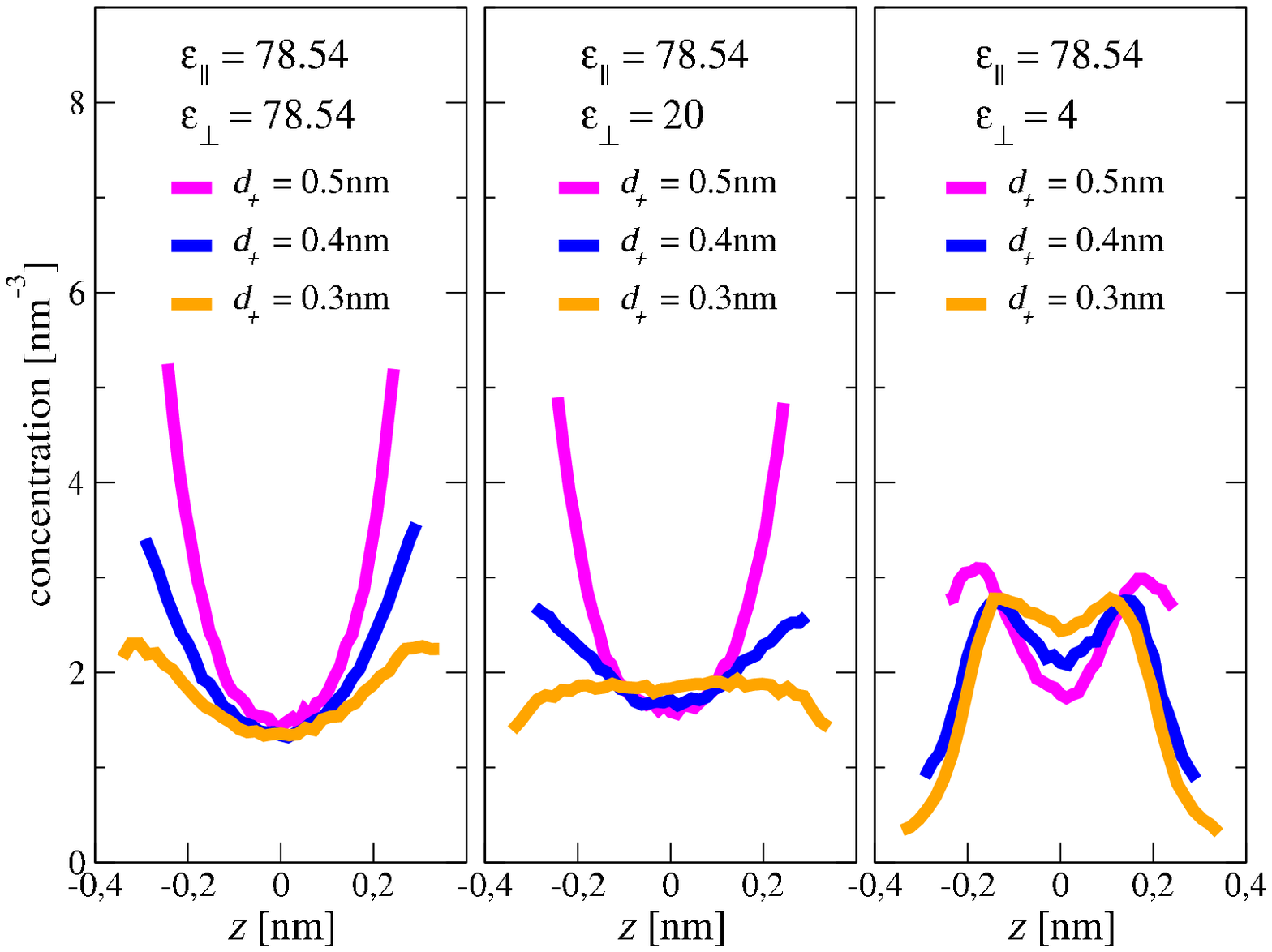}
\caption{Concentration profiles of cations for different diameter $d_+$ and three different dielectric scenarios explained in legends, for dielectric confining surfaces. The anion size is fixed at $d_- = 0.6\text{ nm}$.}
\label{fig:difprofs_diel}
\end{figure}
To clarify how steric asymmetry influences ion distribution, Fig.~\ref{fig:difprofs_diel} shows the equivalent profiles when ionic sizes are altered. When ion sizes are varied, the positions of the respective contact planes shift accordingly. For low values of $\epsilon_{\perp}$, strong in-plane correlation effect dominates: the cations consistently prioritize migrating into the specific plane of closest approach dictated by the hard-core exclusion of the dominant anionic layer, minimizing the overall structural energy of the confined fluid.

\subsection{Ideal Metallic Surfaces}

\begin{figure}[h]
\includegraphics[scale=0.3]{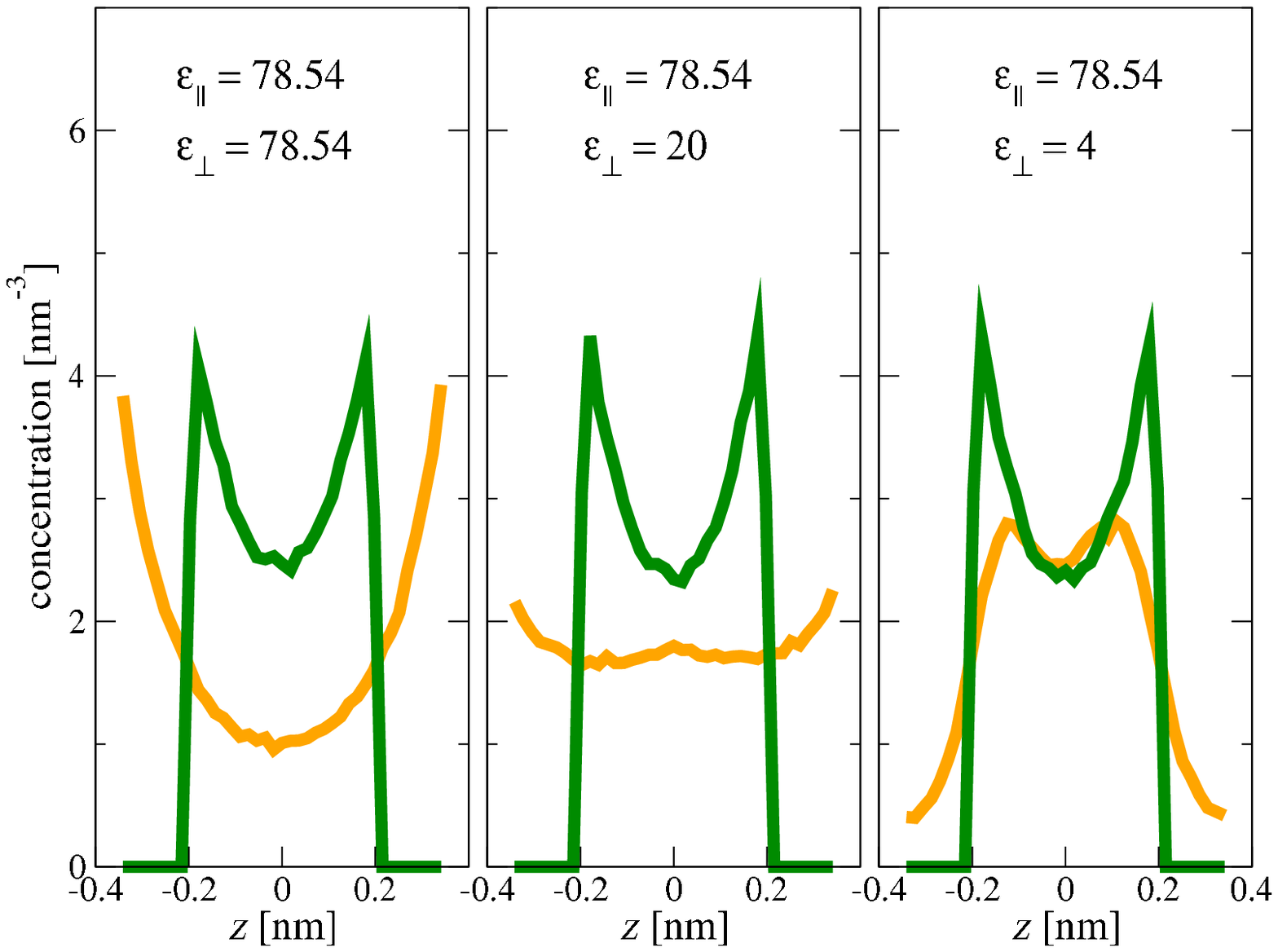}
\caption{Concentration profiles of cations (orange) and anions (green) for three different dielectric scenarios explained in legends, for metallic confining surfaces.}
\label{fig:profs2}
\end{figure}

We next turn our attention to conducting, metallic boundaries where $\epsilon_o = \infty$, yielding a negative contrast factor ($\gamma = -1$) and converting the self-image interaction into an attractive potential. Fig.\ref{fig:profs2} shows the concentration profiles for the asymmetric electrolyte mixture across varying degrees of dielectric anisotropy under these attractive conditions.

In stark contrast to the dielectric walls, the attractive self-image well pulls the ions toward the surfaces, but the spatial distribution remains severely constrained by the ionic hard cores. The large anions maintain an exceptionally stable, highly concentrated monolayer at their contact plane, which is relatively insensitive to variations in the perpendicular permittivity. However, the distribution of the smaller cations exhibits a dramatic sensitivity to the reduction of $\epsilon_{\perp}$. 
Since ion-ion pairing scales with $1/\sqrt{\epsilon_{||}\epsilon_{\perp}}$, the intense in-plane correlations compete directly with attractive image forces and structural packing constraints. As $\epsilon_{\perp}$ drops, the enhanced lateral repulsion between like-charged ions and the enhanced lateral attraction between opposite charges within the dense boundary layers reshape ionic density profiles.

\begin{figure}[h]
\includegraphics[scale=0.3]{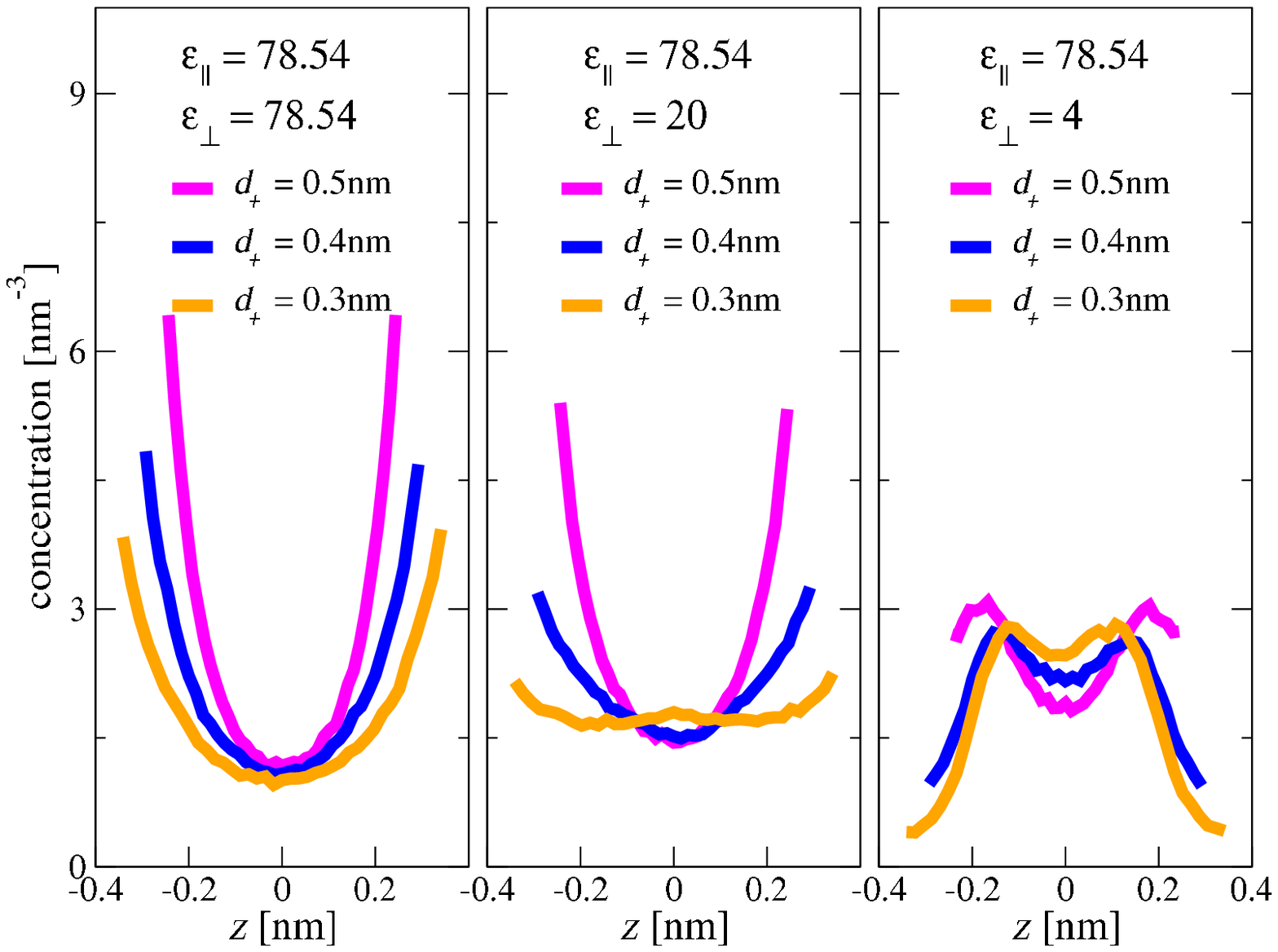}
\caption{Concentration profiles of cations for different diameter $d_+$ and three different dielectric scenarios explained in legends, for metallic confining surfaces. The anion size is fixed at $d_- = 0.6\text{ nm}$.}
\label{fig:difprofs_metal}
\end{figure}
This intricate balance is further detailed in Fig.~\ref{fig:difprofs_metal}, which illustrates the cation density profiles for various ionic size ratios under metallic confinement. While the anionic profiles remain qualitatively anchored due to steric crowding and stable image attraction, the cation distributions display complex shifts in their peak magnitudes and split-peak structures. The suppression of $\epsilon_{\perp}$ effectively compresses the electrical double layer laterally, altering the effective capacitance of the interface. Furthermore, we observe that when $\epsilon_{\perp}=4$, the distinction between metallic and dielectric surfaces completely disappears, and the ionic density profiles for both boundary conditions become practically identical. This clearly demonstrates that for a strongly anisotropic medium, the image charge effects become a minor correction compared to the dominant direct cation-anion correlations, which force cations and anions to preferentially occupy the same $z$-plane. The results presented clearly show that a simple isotropic approximation completely fails to capture the subtle trade-offs between packing geometry and the low-dimensional correlation zones that emerge due to anisotropic dielectric suppression.

\section{Conclusions}

In this work, we have presented a simulation framework to study coarse-grained model electrolytes confined within narrow slit pores characterized by a uniaxial, anisotropic dielectric permittivity tensor. By combining a 2D periodic Green's function approach for the surface polarization fields with a slab-corrected anisotropic 3D Ewald summation, the model accounts for the interplay between steric exclusion, multi-body correlations, and dielectric boundary conditions without requiring the heavy computational burden of explicit-solvent calculations. 

A key aspect of this approach is the precise decoupling of how dielectric anisotropy—specifically the suppression of the perpendicular permittivity component ($\epsilon_{\perp}$) due to the rotational constraints of water in sub-nanometer pores—shapes different components of the electrostatic landscape. By applying a coordinate stretching transformation along the confinement axis, $\tilde{z} = \sqrt{\epsilon_{\parallel}/\epsilon_{\perp}}z$, the system maps onto an isotropic frame governed by an effective dielectric constant $\epsilon_{\text{eff}} = \sqrt{\epsilon_{\parallel}\epsilon_{\perp}}$. Through this mapping, it can be shown analytically that the self-image interaction potential with an isolated planar boundary scales as $U_{\text{self}}(d) = \gamma q^2 / (4 \epsilon_{\parallel} d)$, where the dielectric contrast factor $\gamma$ depends on $\epsilon_{\text{eff}}$. This demonstrates that while $\epsilon_{\perp}$ influences the boundary reflection coefficient $\gamma$, its structural role in the spatial scaling near the wall is entirely canceled by the coordinate deformation. For the specific case of conducting metallic walls ($\epsilon_o = \infty$), $\gamma$ reduces to $-1$, meaning the individual electrostatic well experienced by an isolated ion becomes completely invariant to $\epsilon_{\perp}$ and matches a homogeneous bulk medium.

Crucially, however, the simulations reveal that the suppression of $\epsilon_{\perp}$ reorganizes the electrolyte structure through a modification of the direct ion-ion pair interactions. Because the modified Coulomb law in the stretched coordinate frame scales inversely with $\epsilon_{\text{eff}}$, the lateral interaction between any two ions residing within the same transverse plane ($z_i = z_j$) undergoes a significant amplification. As $\epsilon_{\perp}$ drops from its bulk value of $78.54$ down to $20$ and $4$, the in-plane attraction and repulsion scale as $1/\sqrt{\epsilon_{\parallel}\epsilon_{\perp}}$, driving intense local ionic correlations that force oppositely charged ions to preferentially occupy the same $z$-plane.

This mechanism accounts for the distinct structural variations observed in the simulation profiles under both polarizable dielectric and metallic boundaries. For polarizable dielectric surfaces ($\gamma > 0$), the walls repel the ions, the large anions are sterically driven to pack against their plane of closest approach. Despite the individual image charge barrier, for low $\epsilon_{\perp}$, the amplified in-plane attractive potential completely dominates the double-layer thermodynamics, drawing the smaller cations out of their own preferred positions and locking them directly into the exact same $z$-plane as the anions to optimize local charge neutrality and minimize structural energy.

For conducting walls ($\gamma = -1$) the image interaction becomes attractive. While the large anions continue to form an exceptionally stable, highly concentrated monolayer at their contact plane, the distribution of the smaller cations displays a highly non-monotonic sensitivity to both ionic size ratios and the degree of dielectric anisotropy. Notably, when $\epsilon_{\perp}=4$, the distinction between metallic and dielectric surfaces completely disappears, and the ionic density profiles for both boundary conditions become practically identical, illustrating that the self-image forces become a minor correction compared to the dominant direct in-plane ion-ion correlations.

In summary, these findings indicate that treating nano-confined aqueous environments using a simple isotropic bulk dielectric constant fails to capture the fundamental physics governing double-layer electrostatic assembly. The anisotropic suppression of $\epsilon_{\perp}$ alters the internal fields, generating cooperative, low-dimensional ionic correlation zones within the pore. The generalized methodology and mechanisms established in this work provide a robust framework to explore these phenomena further, offering predictive insight for the refinement of nanofluidic channels, supercapacitors, and electrochemical energy storage devices.

\section{Acknowledgments}

This work was supported by CNPq under the grant 303310/2025-1.

\bibliography{ref.bib}

\end{document}